\documentstyle[12pt,epsf]{article}
\def\refm#1{(\ref{#1})}
\def\beq{\begin{equation}}
\def\h{\frac12}
\def\en{\end{equation}}
\input psfig
\begin{document}
\title {
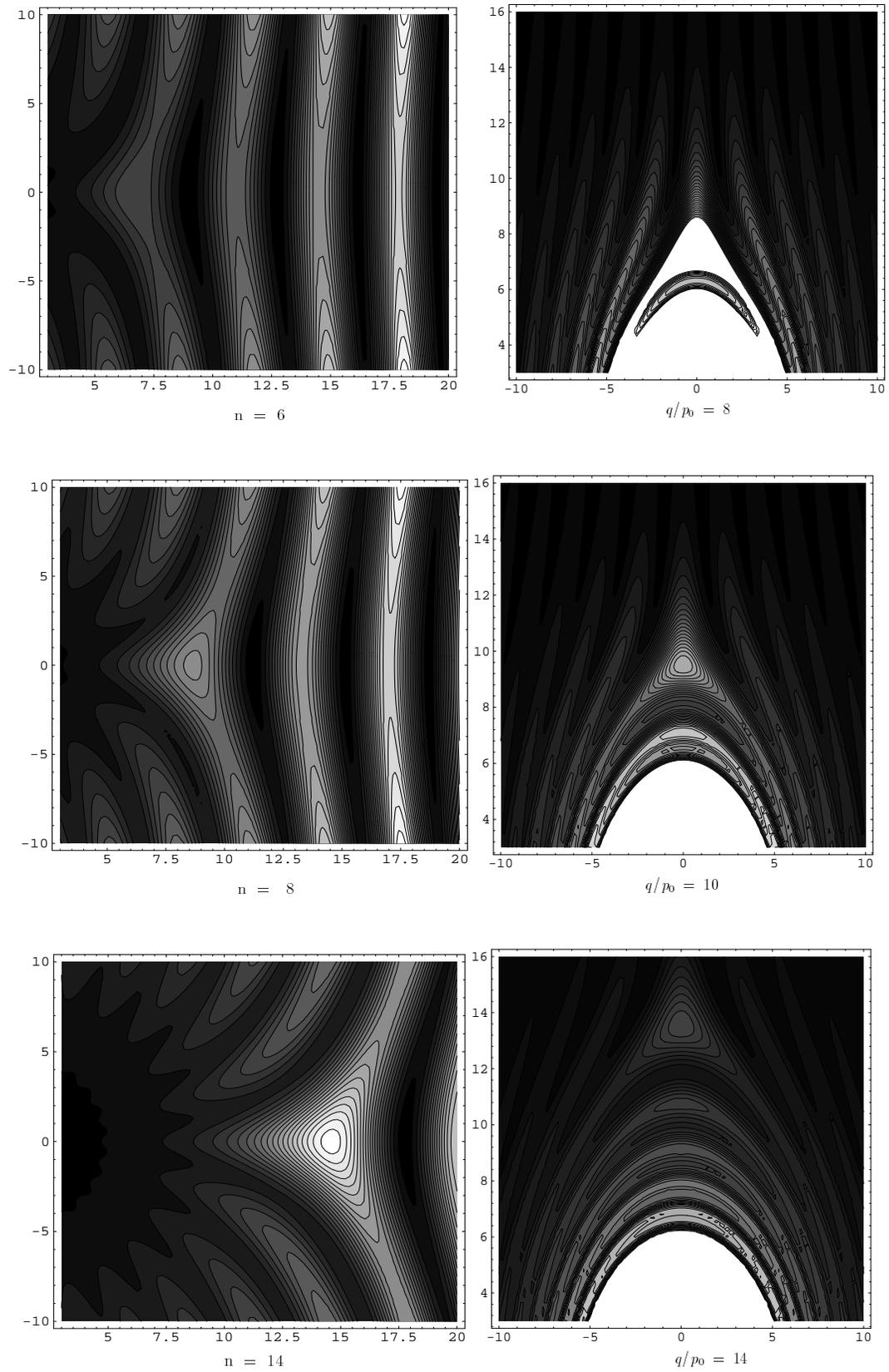
Small-angle scattering on a system of magnetic and
electric charges}
\author{M. Penttinen$^\diamond$ \\
Institut f\"ur Theoretische Physik II \\
Ruhr--Universit\"at Bochum \\
44780 Bochum, Germany}
\maketitle

\begin{abstract}
The generalization of the conformal scattering method for
small-angle scattering processes involving magnetic monopoles
and ordinary charges is constructed.  Using this generalization
we show that introducing of magnetic charges corresponds to
analytical continuation of the eikonal amplitude in the complex
charge plane (the imaginary part is proportional to the magnetic
charge).  We calculate explicitly the eikonal amplitude for
scattering on a dyon and two monopoles in terms of confluent
hypergeometric functions.  The singularities of the
corresponding amplitudes (focal points) are studied in details.
\end{abstract}

\vspace{0.5cm}

\noindent
{\small
$^\diamond$DAAD fellow,
on leave of absence from St. Petersburg University,
St. Petersburg, Russia }

\section{Introduction}
It is known
\cite{Demkov},
that if the potential energy satisfies the Laplace equation, the
small-angle scattering is closely related to conformal
mapping  on a complex plane.
This makes it possible to introduce complex
variables  in the impact-parameter plane $ (b) $
and in the momentum-transfer plane $ (p) $, which leads to a remarkable
simplification of the calculations.
Besides the simplification for harmonic
scatterers, it is possible to formulate certain rigorous theorems
that permit to estimate the classical scattering at  small and
large momentum transfers, as well as to elucidate the nature of the
singularities in the differential effective cross section.

In the present paper we extend the method of conformal scattering
\cite{Demkov,Abramov_Demkov}
to the case of scattering processes involving magnetic monopoles (dyons)
and ordinary charges.

The hypothesis of a new particle, the magnetic monopole, has been introduced
by Dirac \cite{Dirac} in 1931.
Such a particle, for which there is no experimental
evidence would be a source of Coulomb-like  magnetic field.
Dirac argued that in order to fit into conventional quantum
mechanics, the neu particle  must satisfy a quantization condition:
\beq
eg \, =\, n/2
\label{quantcond}
\en
where $g$ is the charge of the monopole, $e$ is the electric charge of any other particle, and $n$ is some integer.
Due to the quantization condition \refm{quantcond} a perturbative
treatment of the
scattering processes involving magnetic monopoles and usual charges is doubtful
if at all possible. In this situation one looks for methods of calculating the scattering amplitude which although approximate, are nonperturbative.

The eikonal approximation \cite{Landau_Lifshitz} for small angle scattering
of fast particles not only makes it possible to express the scattering amplitude
as an integral, but also preserves many important properties of the exact
solution. For
example, it preserves the unitarity relation, which is violated by e.g. the Born approximation. In limiting cases the eikonal approximation reduces to the Born or classical approximations for small-angle scattering.

The small-angle scattering problem simplifies significantly in the case of  harmonic scatterers ( i.e. when the  potential energy satisfies the Laplace equation)
due to the analyticity of complex impact-parametr plane mapping to momentum-transfer plane
\cite {Demkov,Abramov_Demkov}.
After the complex variables, namely the impact parameter
$b$ and the momentum transfer  $p$, are introduced the integrations in the
eikonal formula can be
separated and therefore the scattering
 amplitude factorizes, i.e., it can be expressed as a sum each term of
which is the product of two functions depending on $p$ and $p^*$,
respectively. Thus the scattering amplitude can be expressed through contour integrals,
what  is very
convenient  for both  general studies and numerical calculations.

The outline of the paper is as follows.
In section 2 we formulate the method of conformal scattering for processes involving
magnetic and electric charges. We show that the description of small-angle
scattering processes
on a system of electric and magnetic charges can be simplified by introducing the complex  charge planes (the electric charge corresponds to the real part and
the magnetic charge to the imaginary part).
This  implies that the amplitudes for scattering on various configurations of electric and magnetic charges are related  by analytical continuation of the
amplitudes in the complex charge plane.
In section 3 the quantum mechanical generalization of the method of conformal
scattering is considered. Using this method we examine in details scattering on
a dyon and two monopoles  located at different points.
For the first time we obtain explicit expressions for the eikonal amplitude
for all these problems in terms of confluent hypergeometric functions and Bessel functions of various types
for all these problems.

In section 4 we investigate the focal points, where the classical scattering
amplitude becomes infinite, which play an important role in the general theory of
diffraction and in catastrophe theory.

\section{Classical conformal scattering }

To begin with, let us consider according to
\cite{Demkov} the classical small-angles scattering
of particles  by nonspherically symmetrical scatterer with
potential energy
$U(x,y,z)$ in the presence of an external magnetic field with potential $ A_{i}(x,y,z)$. It can be regarded as a one-to-one single-valued mapping of the
impact-parameter plane $b(x,y)$ into the transverse momentum-transfer
plane $p_\perp(p_x,p_y)$, given by
\[
p_{\perp}= -p_0^{-1}
\nabla V(x,y),
\label{peredannyj_impuls}
\]
\[
V(x,y)= \int_{-\infty}^{\infty} U(x,y,z) dz  - \frac{ep_0}{c} \int_{-\infty}^{\infty} A_z (x,y,z) dz,
\label{potential}
\]
where $e$ and $p_0$ are the charge and the momentum of the incident particle, and
$A_x, A_y \to 0 $ as $z \to \infty $ has been assumed.
The differential effective small-angle cross section is defined as the
Jacobian of the transformation $b \rightarrow  p_{\perp}$
\[
\sigma(p_x,p_y) = p_0^2\sum{\biggl|\frac{\partial(x,y)}{\partial(p_x,p_y)}
\biggr|},
\]
where the summation is over all the values of the multiply valued function
$b(p_\bot)$ (it can be ``zero-valued'' for certain $p_\bot$ ). In other
words, the effective cross section is defined as the ratio of the area
elements in the plane $b$ and the corresponding area elements
 in the plane $p_\bot$.

In the general case, a small circular area in the $b$ plane is mapped
onto an elliptic area in the  $p_\bot$ plane. The line along which
the minor semi-axis of this ellipses vanishes is called the rainbow
line (then the major semi-axes are the tangents of the rainbow line).
The mapping of the rainbow lines on the impact-parameter
plane is defined by the equation
\beq
\partial(p_x,p_y)/\partial(x,y) = 0.
\label{mapping}
\en
For a spherically symmetrical potential the scattering is
axisymmetric and the rainbow lines are circles in the impact-parameter
plane and in the momentum-transfer plane with the center at the
origin.
If the potential  $V(x,y)$\, is a harmonic function, i.e., it satisfies
 the two-dimensional  Laplace equation
$
\Delta V(x,y)=0
$
then rainbow lines degenerate into
points (the so called focal points), since  equation \refm{mapping} being
supplemented by $ \Delta V(x,y)=0 $ takes the form:
\[
\biggl(\frac{\partial^2V}{\partial x^2}\biggr)^2 + \biggl(\frac{\partial^2V}
{\partial x \partial y}\biggr)^2 = 0.
\]
Hence instead of one  condition that defines the rainbow line on a plane,
we obtain thus two conditions that define in the general case only
points on the impact - parameter plane and on the momentum - transfer plane
(focal points).

It is convinient to introduce
complex  impact parameter and complex transverse momentum
\[
b = x +  i y,\qquad  p = p_x + i p_y
\]
as well as a complex potential analytical function, such that
$ V(x,y) \, = \, \mbox{ Re } V(b) $. We then obtain the following
formulas for the momentum transfer, the focal point, and the effective
cross section
\beq
p^*=-p^{-1}_0\frac{dV}{db},
\quad \frac{d^{2}V}{dbdb^*}=0,
\quad \sigma=p^2_0\sum\biggl|{\frac{db}{dp^*}}\biggr|^2,
\label{classical}
\en
where the summation is over all the values of the multiply valued
function $ db/dp^* $.

Let us consider a system of $N$ point-like electric charges $q_i$ situated at points with coordinates $ \vec{w_i} \, (i=1,..,N)$ and of $M$ point like magnetic charges $g_i$
at the points $\vec{z} _i  \, (i=1,..,N)$.
\footnote{The case when $\vec{z} _i = \vec{w_j} $ for some $i$ and $j$ corresponds to a dyon with charges $(q_i,g_i)$ at the point $\vec{z} _i = \vec{w_j} $}
For such system:
\beq
U(x,y,z) \, =\, \sum_{i=1}^{N} \,  \frac{e q_i}{|\vec{r} - \vec{w_i} |},
\label{poten}
\en

\beq
\vec{A} (x,y,z) \, =\, \sum_{j=1}^{M} \, g_j \cdot
\frac{ [(\vec{n_j} \times ( \vec{r} - \vec{z_j})] \cdot  (\vec{n} \cdot ( \vec{r} - \vec{z_j}))}
{|\vec{r} - \vec{z_j} |(( \vec{r} - \vec{z_j} )^2 -
(\vec{n} \cdot ( \vec{r} - \vec{z_j}))^2 )}.
\label{potmag}
\en
We use a form of the vector potential of magnetic monopole suggested by Dirac
\cite{Dirac}
\beq
\vec{A} (z) \, =\, g  \cdot
\frac{ [\vec{n} \times  \vec{r} ] \cdot  (\vec{n} \cdot  \vec{r} )}
{|\vec{r} |( r^2  - (\vec{n} \cdot \vec{r} )^2 )}.
\label{A_dir}
\en
Here $\vec{n}$ is a unit vector in the direction of the  ``Dirac  string'' being  a line of singularity of the potential \refm{A_dir}.
The two-dimensional potential corresponding to \refm{poten} and \refm{potmag}
has the form
\beq
V(x,y)\, =\, V_{el} (x,y) + V_{mag} (x,y)
\en
where
\beq
V_{el}(x,y)\, =\,  \int_{-\infty}^{\infty} U(x,y,z) dz  \, =\,
\sum_{i=1}^{N} q_i \ln ( \, |\vec{b} -  \vec{b_i} |^2)
\en
with $ b_i \equiv (w_{x_i},w_{y_i}) $ being position of the electric charges in
a impact parameter plane, $ \vec{b} =(x,y)$, and
\begin{eqnarray}
V_{mag}(x,y) \, =\, - \frac{ep_0}{c} \int_{-\infty}^{\infty} A_{z} (x,y,z) \,dz =
\nonumber \\
\, = \,  - \frac{ep_0}{c} \sum_{j=1}^{M} g_i \cdot \mbox{ arctan }
\biggl( \frac{(\vec{b} -  \vec{b_j}) \cdot \vec{n_j} }
{[(\vec{b} -  \vec{b_j}) \times \vec{n_j} ]_z}
\biggr).
\end{eqnarray}
Here  $ \vec{b_j} = (z_{x_j},z_{y_j}) $ are the positions of the magnetic charges, and we choose $n_{j z} =0 $ to satisfy a conditions $A_x, A_y \to 0$
as $z \to \infty $. It is easy  to check that  $V(x,y)$ is a harmonic function.
Introducing a complex impact-parameter plane and a complex potential plane
we see that
\begin{eqnarray*}
V_{el} (x,y)  & = &   \mbox{Re }  \sum_{i=1}^{N} \, q_i \ln (b-b_i)
\\
V_{mag} (x,y) & = & - \frac{qp_0}{c}  \, \mbox{Im }  \sum_{i=1}^{N} \, e g_i
\ln [(b-b_i) \cdot e^{i \theta _j} ]
\end{eqnarray*}
with $\theta _j = \mbox{ arctan } (n_{x_j} /n_{y_j} ) $. The
effective  complex potential for a system of $N$ electric and $M$ magnetic
charges has the form:
\beq
V(b)\, =\, V_{el} (b) + V_{mag} (b)\, =\,\sum_{k=1}^{M+N} \, Q_k
\ln [(b-b_k) \cdot e^{i \theta _k}  ],
\label{V_tot}
\en
where $ Q_k \, =\, e q_k - \frac{iep_0}{c} g_k $ are complex charges with real
part equal to the electric charge $q_k$ at point $b_k$, and with imaginary
charge part proportional to the magnetic charge at the same point. We see that
introducing
magnetic charges into a system corresponds to continuation of the charges into a
complex plane, this fact reduces the problem of scattering on a system
involving
magnetic charges to the corresponding problem with only electric charges by
analytical continuation in the complex charge plane. It is also worth noting that the imaginary part of the complex charges is proportional to the  momentum of
the incident particle.

The general formulas \refm{classical} and \refm{V_tot} being applied  to the
case of scattering on a point-like dyon with charges $q$ and $g$ give a
 Rutherford formula
\beq
\sigma \, =\, \frac{1}{|p|^4} \cdot \bigg\lbrace  \biggl(  \frac{2p_0}{c} eg      \biggr)^2 + (eq)^2 \bigg\rbrace .
\label{Rutherford}
\en

As a next example let us consider the problem of scattering by two dyons
with complex charges $ Q_{1,2} \, =\, e q_{1,2} + ie g_{1,2} p_0 /c $
at the points $b_1=R, \,  b_2= -R $.
In this case equations  \refm{classical} and \refm{V_tot} give
\beq
p^*  \, =\, \frac{1}{p_0} \cdot \biggl( \frac{Q_{1}}{b-R} +
\frac{Q_{2}}{b+R} \biggr).
\en
The equation for the focal points reads:
\beq
\frac{Q_{1}}{(b_f -R)^2} +
\frac{Q_{2}}{(b_f +R)^2}  \, =\, 0
\en
and has the following solutions:
\beq
b_f^{1,2} \, =\, -R \cdot
\frac{( \sqrt{Q_{1}}  \pm i \sqrt{Q_{2}})^2}
{Q_{1} +  Q_{2}}.
\en
The corresponding values of the momentum transfer,
where the cross section is infinitely large, are
\beq
p^f_{1,2} \, =\, - \frac{1}{4Rp_0} \cdot \frac{(Q_{1} +  Q_{2} )^2}
{(\sqrt{Q_{1}} \pm i \sqrt{Q_{2}})^2}
\en
and the cross section can be written in the form:
\beq
\sigma  \, =\,  \frac{| Q_{1} +  Q_{2} |^2}{2|p|^4} \cdot
\biggl[ \,
1+ \frac{1}{4} \biggl| \, 2 \, + \frac{p/p_1^f -1}{p/p_2^f -1} +
\frac{p/p_2^f -1}{p/p_1^f -1} \,  \biggr| \,
\biggr].
\en
It has a form similar to the one of the cross section for scattering on two Coulomb centres
\cite{Demkov}. For equally charged monopoles $ eg_1 =  eg_2 = n/2 $ we have
\[
p^f_{1,2} \, =\, \pm n/R
\]
and
\beq
\sigma  \, =\,  \frac{n^2 p_0^2}{2|p|^4} \cdot
\biggl( \,
1+  \biggl| \, \frac{1}{1-  p^2 \cdot R^2 /n^2 }  \,  \biggr| \,
\biggr).
\en
For monopole-antimonopole pair with  $ eg_1 = - eg_2 = n/2 $ one obtains:
\beq
\sigma  \, =\,  \frac{n R p_0^2}{2|p|^3} \cdot
\frac{1}{ | \, 1-  iRp /n  \, |}, \quad  p_1^f=0, \quad
p_2^f = - \frac{in}{R}.
\en
In the limit of small $|p| \ll \frac{n}{R}$ one gets a scattering on magnetic dipole with moment $d  \, =\, nR/e $.

\section{ Quantum conformal scattering}

Let us now consider scattering on a system of electric and magnetic
charges in the eikonal approximation.
As mentioned above the theory becomes greatly simplified for
harmonic potentials \cite{Abramov_Demkov}.
In the eikonal approximation the scattering amplitude can be expressed
as a double integral over the impact-parameter plane
\beq
f = \frac{p_0}{2\pi i}\int\limits_{-\infty}^{\infty}dx\int
\limits_{-\infty e^{ i\epsilon}}^{\infty e^{- i\epsilon}}
dy \exp{\biggl[- i(xp_x +yp_y) + \frac{ i}{p_0}V(x,y)\biggr]},
\label{eikonal_amplituda}
\en
where
\beq
V(x,y)\, =\, \int\limits_{-\infty}^{\infty} dz U(x,y,z) - \frac{ep_0}{c}
\int\limits_{-\infty}^{\infty} dz A_z(x,y,z)
\en
is, as previously, a harmonic function.
To pass to the classical limit one may assume the argument of the
exponential in \refm{eikonal_amplituda} to be large and evaluate
the integral by the saddle point method.

The potential $ V(x,y) $ is a harmonic function, and hence it  can be represented as a real part
of analytical function of
the variables   $b = x + i y $ (the complex impact parameter) and
$b^* = x- i y $, where $x$ and $y$ are real:
\beq
V(x,y)= \biggl[ V(b)+V^*(b^*)\biggr].
\en
It was shown in \cite{Abramov_Demkov} that passing  from variables
\,$x$\, and \,$y$\,
to \,$b$\, and \,$b^*$\,
leads to factorization of the integrand and to the separation of variables
$b$ and $b^*$ which enables us to express the result as product of
contour integrals
\beq
f = \frac{p_0}{4\pi} \,  \sum_{j=1}^{M+N}  I_j(p)\, I_j(p^*),
\label{factor_amplituda}
\en
where $ I(p), I(p^*) $ are functions of the complex momentum
$p=p_x + i p_y $ and of the conjugate momentum
$p*=p_x -  i p_y $,  respectevely, and the sum goes over positions of charges
(electric and magnetic):
\beq
I(p)\, I(p^*) = \int_{C_{b_j^*}}\exp\biggl[-\h i b^* p +\frac{ i}{p_0}
V^*(b^*)\biggr]db^*\int_{C_{b_j}}\exp\biggl[-\h i bp^* +
\frac{ i}{p_0} V(b)\biggr]db.
\label{Ip_Ip*}
\en
The cuts in the complex impact-parameter plane $(b)$ go parallel to the real
axis from $ - \infty $ to each of the points $b_j$ and  $b_j^*$
(positions of charges).
The form of contours $ C_{b_j^*}$ and $ C_{b_j} $ are shown in Fig.1.

Hence for the problem of small-angle scattering by a system of electric and
magnetic charges the scattering amplitude is expressed as
a product of two similar contour integrals,
one of them depending on $p$ and the other on $p^*$.

\subsection{ Special cases }

Let us apply the technique of conformal scattering to some special cases.
First we consider a single dyon with charges $(q,g)$ located at $b_0=x_0+iy_0$
and a Dirac string characterized by the angle $\theta $. The
evaluation of contour integrals $I(p)$ and $I(p^*)$ given by equation
\refm{Ip_Ip*}
with
\[
V(b)\, =\, \biggl( eq +i \frac{egp_0}{c}  \biggr) \ln [ (b-b_0) \, e^{i \theta }],
\]
gives the eikonal amplitude for scattering on a dyon:
\begin{eqnarray}
f(p)\, =\, 2i\, \frac{p_0}{|p|^2} \,
{ \biggl( \sqrt{ \frac{p}{p^*} } \biggr) }^{eg/c}
\biggl( \frac{2p_0}{|p|} \biggr) ^{2iq/p_0 }
\biggl( \frac{iq}{p_0} + \frac{eg}{c} \biggr)
\times
\nonumber \\
\times  \exp (-i \mbox{Re } (b_0 p^*)- i \pi eg/c + 2i eg \theta ) \quad
\frac{\Gamma \biggl( \frac{iq}{p_0} + \frac{eg}{c} \biggr)}
{\Gamma \biggl( \frac{iq}{p_0} - \frac{eg}{c} \biggr)}.
\label{dyon}
\end{eqnarray}
We see that the dependence on the direction of the Dirac string $ \theta $ enters only
as a constant phase and hence it does not affect  any physical observables. It is easy to show that this simple dependence on the Dirac-strings directions is
also pertinent for more general cases of many monopoles. In the case of a system
containing
$M$ Dirac monopoles with charges $g_i \, (i=1, ..., M) $ and ``Dirac-strings
angles'' $\theta _i \, (i=1, ..., M) $ the dependence on the latter has a
simple form of a constant phase:
\[
e^{i \sum_{i=1}^{M}  2eg_i \theta_i /c }.
\]
Using this fact one can set  all $\theta_i = 0$ in all our calculations.
The expression \refm{dyon} for small-angle scattering amplitude on a dyon
coincides with the result of rather tedious calculations
of ref \cite{Calucci}.
The corresponding effective  cross section on the dyon takes the following
form, coinciding with the
classical cross  section \refm{Rutherford}
\beq
\sigma \, =\, \frac{4}{|p|^4} \cdot  \bigg \lbrace
q^2 + \biggl( \frac{eg p_0}{c}\biggr)^2 \bigg \rbrace .
\en
In the special case of a Dirac monopole $e \cdot g = 1/2 $  placed at
$ b=0 $ we obtain the small-angle scattering amplitude:\begin{eqnarray}
f\, =\, - \frac{n}{\pi} \, \frac{p_0}{|p|^2} \,
\biggl( \frac{p}{p^*} \biggr)^{n}
\biggl( \frac{\tau^* }{\tau } \biggr)^ {\frac{n}{2}}
\mbox{\rm exp} \biggl[ -\frac{i}{4} \mbox{\rm Re } 
\biggl( (b_2 + b_1 )p^*\biggr) 
\biggr]
\cdot
\nonumber \\
\cdot
\biggl[
K_{\frac{1+n}{2} } \biggl( \frac{\tau^*}{2} \biggr)
\cdot K_{\frac{1-n}{2}} \biggl( -\frac{\tau }{2} \biggr)
\, -\, K_{\frac{1+n}{2}} \biggl( -\frac{\tau^*}{2} \biggr)
\cdot K_{\frac{1-n}{2}} \biggl( \frac{\tau}{2} \bigg)
\biggr].
\end{eqnarray}
\beq
f(p)\, =\, \frac{p_0}{|p|^2} \, \sqrt{\frac{p}{p^*}}
{ \biggl( \frac{2p_0}{|p|} \biggr) }^{2iq/p_0 }
\biggl( 1 + \frac{2iq}{p_0} \biggr) \,
\frac{\Gamma \biggl( \h + \frac{iq}{p_0} \biggr)}
{\Gamma \biggl( \h  - \frac{iq}{p_0} \biggr)}.
\en
Now one can relatively easy  calculate the amplitude for scattering
on two monopoles in the eikonal approximation. Let the complex numbers $b_1$
and $b_2$ be  the projections of the magnetic charges $g_1$ and $g_2$
onto the impact-parameter plane.
Introducing the notations:
\begin{eqnarray*}
 \ae & = &\frac{e}{2} (g_2 -g_1),
\\
 \mu & = &\h - \frac{e}{2} (g_2 +g_1), \qquad \mu ' =1- \mu,
\\
\tau & = &\frac{ip}{2} (b_2 - b_1)^* , \,
\end{eqnarray*}
after simple calculations we obtain the following expressions for integrals
\refm{Ip_Ip*} in the terms of the Whittaker function
\cite{Erdelyi} $ W_{\ae , \mu}$:

\begin{eqnarray*}
I(p) & = & 2\pi i\,(-1)^{- \ae }
e^{ - \frac{1}{4}ip (b_1^* +b_2^*)}
{(b_1^* - b_2^*)}^{2\mu} \,
\tau ^{- \mu -\h}
\frac{1}{\Gamma \biggl( \ae - \mu + \h \biggr) }W_{ \ae , \mu  } (-\tau),
\\
I(p^*)& = & e^{ - \frac{1}{4}ip^* (b_1 +b_2) }  {(b_1 - b_2)}^{2 \mu'} \,
{\tau ^* }^{- \mu ' -\h} \,
\Gamma \biggl( - \ae  + \mu ' + \h \biggr)
W_{- \ae , \mu ' } (\tau ^*).
\end{eqnarray*}

Implying Dirac's quantization conditions  $eg_1 \, =\, n_1/2 $,  $eg_2 \, =\, n_2/2 $,
where \mbox{ $ (n_1,n_2)  \in \bf{Z} $ }, one gets the
amplitude for scattering on two monopoles:

\begin{eqnarray}
f(p)\, =\,-  \frac{i p_0}{|p|^2} \,
{\biggl( - \frac{p^*}{p}\biggr)}^{\frac{n_1 + n_2}{2}}
{\biggl(  \frac{\tau  }{\tau ^*} \biggr) }^{\frac{n_1 + n_2}{4}}
\, \exp \bigg \lbrace - \frac{i}{4} \mbox{ Re }((b_1 +b_2)p^*) \bigg \rbrace
\times
\nonumber \\
\times
\bigg \lbrace \,
(-1)^{n_1/2- \frac{n_1+n_2}{4} } \cdot n_1
W_{ \frac{n_1 - n_2}{4} , \h - \frac{n_2 + n_1}{4}}(\tau ^*) \cdot
W_{  \frac{n_2 - n_1}{4} , \h + \frac{n_2 + n_1}{4}}(-\tau ) +
\nonumber \\
+\, (-1)^{n_2/2+ \frac{n_1+n_2}{4} }\cdot n_2 
W_{ -\frac{n_1 - n_2}{4} , \h - \frac{n_2 + n_1}{4}}(- \tau ^*) \cdot
W_{ - \frac{n_2 - n_1}{4} , \h + \frac{n_2 + n_1}{4}}(\tau )
\bigg \rbrace .
\end{eqnarray}

The formula for the eikonal amplitude for the scattering by two identical
monopoles $ n_1 =n_2=n$ simplifies, in this is case
$ \ae = 0 $ and hence we can use the representations for the Whittaker functions
in terms of modified Bessel function  $K_{\mu }(z)$:
\[
W_{ 0, \mu }(z)\, =\, \sqrt{\frac{z}{\pi} }K_{\mu } \biggl( \frac{z}{2}
 \biggr).
\]
Using this relation one obtains the scattering amplitude for small-angle scattering on two identical monopoles with charges $eg=n/2$:
\begin{eqnarray}
f\, =\,  \frac{n}{2 \pi} \, \frac{p_0 |b_2-b_1|}{|p|} \,
\biggl( - \frac{p^*}{p} \biggr)^{n}
\biggl( \frac{\tau^* }{\tau } \biggr)^ {\frac{n}{2}}
\mbox{\rm exp} \bigg \lbrace  -\frac{i}{4} \mbox{\rm Re } 
((b_2 + b_1 )p^*) 
\bigg \rbrace
\times
\nonumber \\
\times
\biggl[
K_{\frac{1-n}{2} } \biggl( \frac{\tau^*}{2} \biggr)
\cdot K_{\frac{1+n}{2}} \biggl( -\frac{\tau }{2} \biggr)
\, -\,(-1)^n  K_{\frac{1-n}{2}} \biggl( -\frac{\tau^*}{2} \biggr)
\cdot K_{\frac{1+n}{2}} \biggl( \frac{\tau}{2} \bigg)
\biggr].
\end{eqnarray}
The amplitude for the case of small-angle scattering on $M$ monopoles is
expressed in terms of generalized hypergeometric functions.

\section{Focal points}

The small angle classical cross section for scattering of a charge
particle on a system of point-like electric and magnetic charges diverges
at particular values of the momentum transfer (focal points $p_f$) determined by eq.
\refm{mapping}. In the
quantum case this  divergence is  shadowed by quantum corrections, but in the
semiclassical limit
the amplitude has clear  maxima in a neighbourhood of focal points.
The semiclassical amplitude in the neighbourhood of the focal points is not given
by the steepest descent method for integral \refm{eikonal_amplituda}.
For spherically symmetric
scatterers the  semiclassical amplitude in the neighbourhood of the focal points
is given by the  Airy function \cite{Newton}. The more complicated case of
two-centre
Coulomb scatterer was considered in \cite{Abramov_Demkov}. It was shown
there that the corresponding amplitude in the neighbourhood of the focal points
is described  by canonical integral, which is called  elliptic
umbilica \cite{Berry} in catastrophe theory.

We consider in details the approach to the semiclassical limit of the amplitudes
in the neighbourhood of a focal point for two cases: scattering on two Coulomb centres
and on two monopoles.
Let us consider two systems: the first one with  two equal electric charges $q$
placed and $b= \pm R$, the second one is a system of two equal magnetic charges with
\mbox{ $ eg=n/2 \,(n \in \mbox{\bf Z})$}.
The values of the momentum transfer corresponding to focal points are given for
the first system \cite{Abramov_Demkov} by \mbox{$p_f= \pm 2iq/ p_0 R $} and for
the second one by $p_f= \pm n/ R $  (see section 2).
The scattering becomes semiclassical if the following conditions are satisfied:
$p_0R \gg 1 $
(incident particles wavelength much less than the  size of the system) and
$\theta_f  = \biggl| \frac{p_f}{p_0} \biggr| \gg 1 $
(classical scattering angle larger than the quantum mechanical
uncertainty of the scattering angle).
The above conditions imply for the first system that  $q/p_0 \gg 1$
and $n \gg 1$ for the second one. In Fig.2 we show the appearance of maxima in cross sections corresponding to focal points for both systems with approaching
to the classical limit at increasing values of $q/p_0$ and $n$
for the first and the second system correspondingly.
\footnote{For electric case we plot $|f|^2$ and for magnetic case
we plot $|pf|^2$ to see focal more clearly.}
The cross sections in the neighbourhood of the focal points are shown in Fig.3,
one sees that for both systems the cross sections have clear maxima
and have a similar behaviour corresponding to elliptic umbilica.


\section{CONCLUSION}

The method of conformal scattering is generalized to the case of systems
involving
magnetic charges besides the electric ones. Calculations of the scattering
amplitudes in the eikonal approximation are simplified  considerably due to
conformal properties of mapping of the impact parameter plane
to the momentum transfer
plane. This property enables us to introduce the notion of two dimensional complex analytical potential. The complex potential for a system involving magnetic charges corresponds to a system of Coulomb centres with complex charges having
imaginary part proportional to the magnetic charge and the momentum of incident
particle. Using these simplifications we obtained an explicit expression for the
small-angle scattering amplitude on a system of two monopoles in terms of
confluent hypergeometric functions for generic case and Bessel functions for
two equal magnetic charges.

These explicit formulas enable us to study the semiclassical limit of the
corresponding amplitudes in the neighbourhood of rainbow singularities (focal points).
The behaviour of the cross-sections in the neighbourhood of these point singularities
has a form of elliptic umbilica catastrophe.

\section{ACKNOWLEDGMENT}

This work has been supported  by Deutscher  Akademischer  Austauschdienst (DAAD).
The author is indebted to Prof. Yu. Demkov for numerous discussions of
method of conformal scattering. The author is sincerely grateful to Prof. K.
Goeke for warm hospitality.

\end{document}